\documentclass[aps,prb,twocolumn,amsmath]{revtex4}
\usepackage{graphicx}
\usepackage{dcolumn}
\usepackage{bm}

\begin{document}

\title{Directional emission of light from a nano-optical Yagi-Uda antenna}

\author{Terukazu Kosako}
\author{Holger F. Hofmann}
\email{hofmann@hiroshima-u.ac.jp}
\author{Yutaka Kadoya}
\email{kd@hiroshima-u.ac.jp}
\affiliation{
Graduate School of Advanced Sciences of Matter, Hiroshima University,
Kagamiyama 1-3-1, Higashi Hiroshima 739-8530, Japan}

\maketitle 

\textbf{The plasmon resonance of metal nanoparticles can enhance and direct light from optical emitters in much the same way that radio frequency (RF) antennas enhance and direct the emission from electrical circuits. 
Recently, rapid progess has been made in the realization of single element antennas for optical waves 
\cite{CrozierJAP03,SchuckPRL05,MuhlschlegelSCI05,KuhnPRL06,SundaramurthyNL06,
AizpuruaPRB05,FrommNL04,FarahaniPRL05,FrommJCP06,MuskensNL07,BakkerAPL08,BakkerNJP08}.
Since most of these devices are designed to optimize the local near field coupling between the antenna and the emitter, the possibility of modifying the spatial emission pattern  has not yet received as much attention \cite{TaminiauNL07,TaminiauNP08}. 
In the RF regime, a typical antenna design for high directivity is the Yagi-Uda antenna, which basically consists of a one-dimensional array of antenna elements driven by a single feed element.
By fabricating a corresponding array of nanoparticles, similar radiation patterns can be obtained in the optical regime \cite{LiPRB07,OursNJP07}.
Here, we present the experimental demonstration of directional light emission from a nano-optical Yagi-Uda antenna composed of an array of appropriately tuned gold nanorods.  
Our results indicate that nano-optical antenna arrays are a simple but efficient tool for the spatial control of light emission. 
}

Directing the fluorescence from nano-scale emitters such as molecules and single quantum dots is a key issue in a wide range of applications from the efficient detection of molecules for biological diagnostics to photonic devices for quantum-information processing.
It has long been known that the fluorescence of quantum emitters can be modified by controlling the optical resonances in the vicinity of the emitter \cite{PurcellPR69}.
Experimentally, the effect has been demonstrated using optical cavities composed of a pair of metallic \cite{HuletPRL85} or dielectric mirrors \cite{BjorkPRA91} or photonic crystals \cite{OgawaSCI04}.
In principle, the plasmon resonances of nano-particles could achieve similar results on even smaller length-scales\cite{FarahaniPRL05,KuhnPRL06,BakkerAPL08,BakkerNJP08,MuskensNL07,TaminiauNP08}, offering a promissing alternative to cavity resonators for use in integrated nano-optical devices.

To develop the full potential of directional emission from nano-particles, we take our inspiration from RF technology, where Yagi-Uda antennas are known to achieve high directivities.
 Yagi-Uda antennas are composed of a linear array of metal rods working as feed, reflector, and directors \cite{YagiUda1}.
The electromagnetic wave emanating from the feed element induces currents in the other passive elements of the antenna array, resulting in phase coherent emissions from all the elements in the array. 
To obtain strong directivity, the phase shifts in the electromagnetic response of the elements must be adjusted so as to result in constructive interference along the forward direction and destructive interference in other directions.
Figure 1 illustrates the typical geometry of a Yagi-Uda antenna.
Since resonant elements cause reflections, the directors in front of the feed element are capacitively detuned to resonate at wavelengths shorter than the emitted wavelength $\lambda$.
A single reflector element inductively detuned to wavelengths longer than $\lambda$ is placed behind the feed to further reduce the radiation emitted in the backward direction. 
Optimal directivity is obtained for distances of around $\lambda/\pi$ between neighbouring elements, except for the distance between reflector and feed, which produces best results around $\lambda/4$.
\begin{figure}[tbp]
    \includegraphics[keepaspectratio=true,width=75mm]{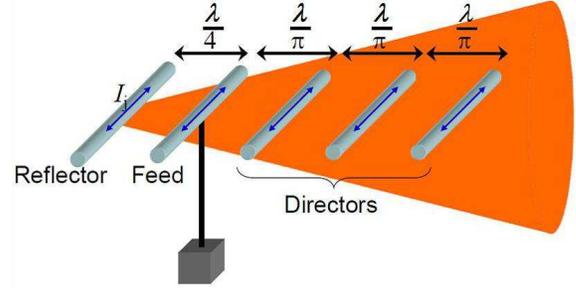}
  \caption{Typical geometry of a 5-element Yagi-Uda antenna.}
  \label{fig:yagi-uda.eps}
\end{figure}

In the radio frequency regime, antenna elements are realized by metal rods resonating at wavelengths of about twice their length.
At optical frequencies, it is necessary to take into account the precise electromagnetic response of the metal as given by its complex dielectric constant.
Specifically, the optical response of metals is characterized by a negative real part that originates from the retardation of electron motion due to the electron mass.
For sufficiently small particles, the kinetic energy of the electrons replaces the magnetic energy of the inductance in the electromagnetic response of the particle, resulting in a size-independent localized surface plasmon resonance (LSPR) determined by the shape and the dielectric constant of the particle.
As a result, the resonant wavelength of a nanoparticle is determined not by its length but by its shape and composition. To realize Yagi-Uda antennas for light waves, it is therefore necessary to tune the resonances of the elements by varying their geometrical structure \cite{LiPRB07,OursNJP07}. 
 \begin{figure*}[htbp]
      \includegraphics[keepaspectratio=true,height=100mm]{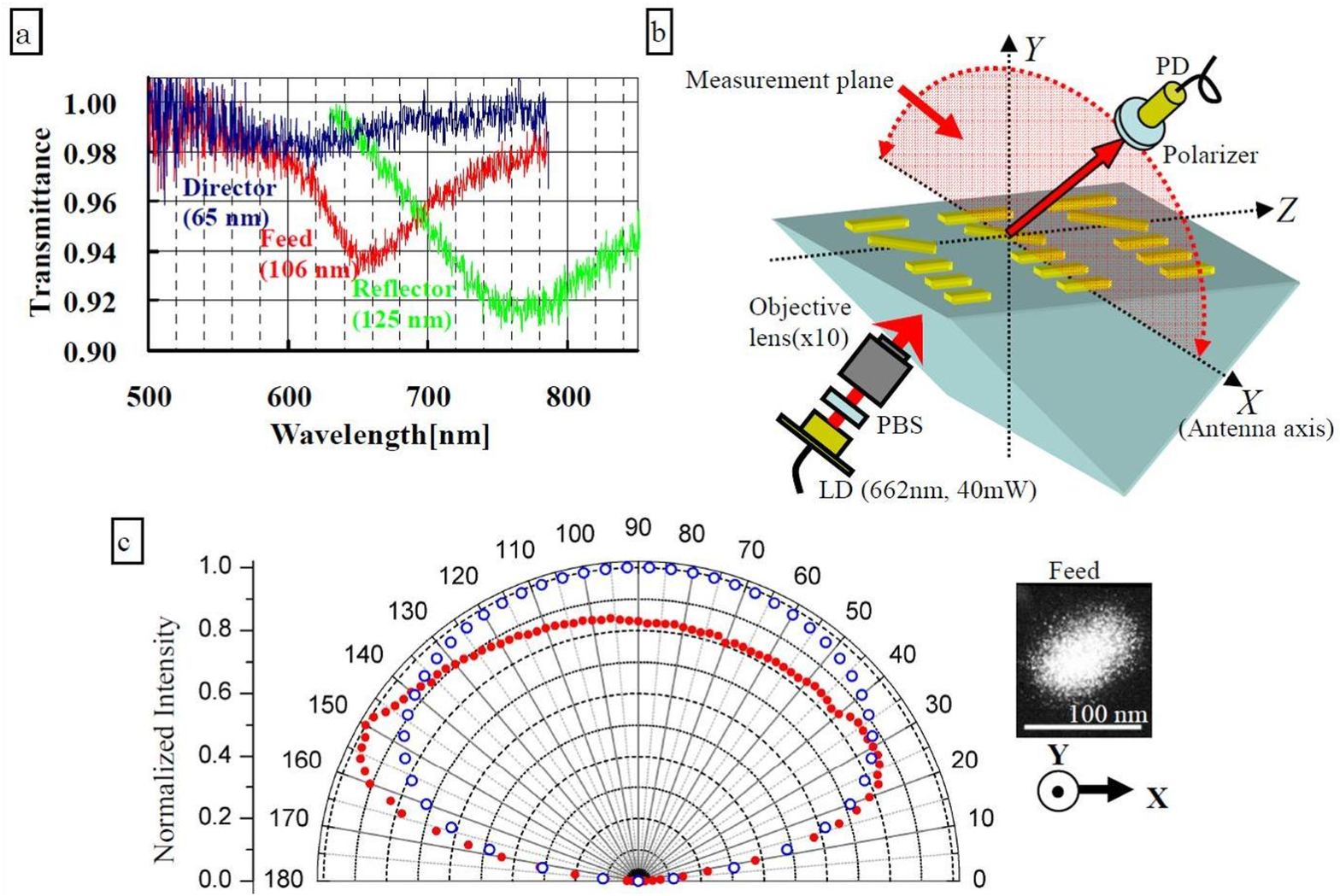}
      \caption{{\bf a}, Transmission spectra for the three different nanorod geometries used in the antenna array. {\bf b}, Measurement set-up for detecting the directionality of the optical antenna. The spherical lens used in the measurement is not shown. {\bf c}, The emission pattern of a feed element. The solid dots and the open circles represent the experimental result and the theoretical prediction, respectively. The inset shows the SEM image of the feed element.}
      \label{fig:fig2.eps}
 \end{figure*}
In the case of nanorods, the resonance depends on the aspect ratios of the rod geometry, with elongated rods resonating at longer wavelengths than spherical particles, regardless of size.
In addition to the negative real part of the dielectric constant, it is also necessary to consider its imaginary part, which describes the absorption of radiation energy in the material.
For antenna elements, it is essential that the losses due to absorption are sufficiently smaller than the emission of radiation into the far field, so that the energy of the plasma oscillation can be emitted before it is turned into heat inside the metal. 
As we have shown in our previous work, the ratio between material losses and radiation losses decreases with the total volume of the particle \cite{OursNJP07}. This means that material losses impose a minimal size on the nanoparticles used in the antenna array. We find that for gold, this size is in the range of 50 nm to 100 nm.

The essential properties of the nanorods can be summarized in terms of their volume $V$, a shape-dependent depolarization factor $N$ \cite{Dipole}, and the dielectric constant $\epsilon_r$  of the rod material relative to the surrounding medium ($\epsilon_r=\epsilon_{\rm rod}/\epsilon_{\rm med}$). For a given wavelength $\lambda$, the polarizability $\alpha$ is then given by \cite{OursNJP07}
\begin{eqnarray}
\alpha=\frac{(\epsilon_r-1)V}{1+N(\epsilon_r-1)-i(\epsilon_r-1)\frac{4\pi^2V}{3\lambda^3}}.
\end{eqnarray}
 The dipole resonance is capacitively (inductively) detuned for $N>1/|{\rm Re}[\epsilon_r]-1|$ 
($N<1/|{\rm Re}[\epsilon_r]-1|$).
As mentioned before, the value of $N$ can be adjusted by selecting the appropriate aspect ratio of the nanorods.
For practical purposes, it is convenient to keep the cross section of the nanorods constant while varying only their length.
Thus, the relation between the detuning and the length of the nanorods appears to be the same as in the radio frequency regime, with shorter (longer) rods having shorter (longer) wavelength resonances.
\begin{figure*}[htbp]
      \includegraphics[keepaspectratio=true,width=95mm]{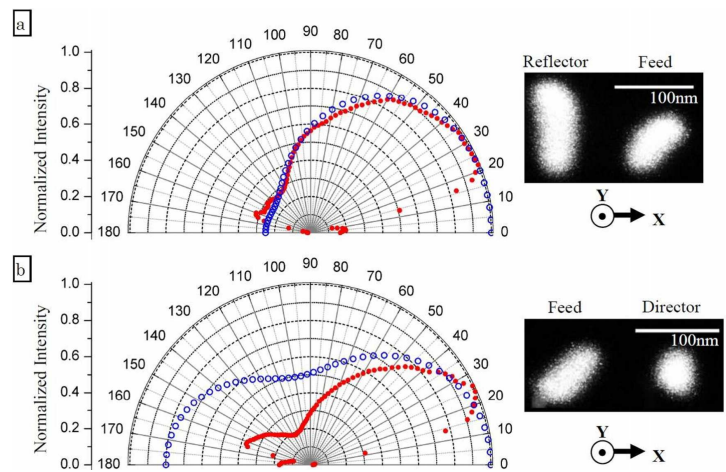}
      \caption{Radiation patterns of two-element antennas. The insets show the SEM images of the antennas. The experimental results and the theoretical predictions are plotted with solid dots and open circles, respectively. {\bf a}, Feed-reflector antenna. {\bf b}, Feed-director antenna.}
      \label{fig:fig3.eps}
 \end{figure*}
\begin{figure*}[htbp]
      \includegraphics[keepaspectratio=true,width=120mm]{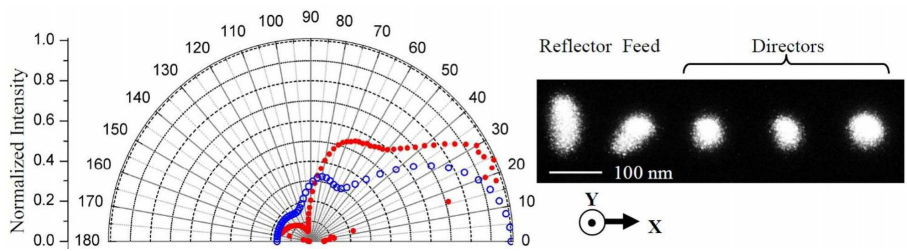}
      \caption{Measured (solid dots) and predicted (open circles) radiation pattern of the five-element Yagi-Uda antenna. The inset shows the SEM image of the antenna.}
      \label{fig:fig4.eps}
 \end{figure*}

For the experiment, antenna arrays made of gold nanorods were fabricated lithographically on a glass substrate. The antennas were then embedded in a sputter-deposited SiO$_x$ film to suppress the effects of a varying index of refractivity on the coupling between array elements.
The resonant wavelength of the different nanorod geometries was confirmed experimentally by measurements of the transmission spectra of light polarized parallel to the major axis of the nanorods. 
Figure 2a shows the transmission spectra for the three different nanorod geometries used in the antenna arrays.
The resonance wavelengths were found to be $\sim$610 nm, 655 nm, and 770 nm for nanorods of length 65 nm (director), 106 nm (feed), and 125 nm (reflector), respectively.

To emulate a local emission process originating from the feed element, we used the strong polarization dependence of the nanorod response to an external driving field. By tilting the feed element 45$^\circ$ toward the antenna axis, we can drive it  with light polarized along the antenna axis without directly exciting the plasma resonances of the passive antenna elements. Due to its diagonal alignment, the feed element converts part of the driving light into light polarized along the major axes of the passive elements, resulting in directed emission along the antenna axis. The directivity of emission originating from oscillations of the feed element can then be detected by measuring the emission of light polarized perpendicularily to the antenna axis.

The feed element was driven by a laser diode at a wavelength of 662 nm, so that the directors are capacitively detuned and the reflectors are inductively detuned.The emitted light was coupled out of the glass medium by a hemispherical lens attached to the sample.The emission in the plane orthogonal to the substrate and including the antenna axis was then scanned with a photo diode detector as illusted in Fig.\ 2b. 
Figure 2c shows the emission pattern of a feed element without any reflectors or directors.
The emission is nearly symmetric around the normal to the surface, as expected for emission from a single dipole.
The slight asymmetry of the pattern is probably caused by a small misalignment between the center of the spherical lense and the antenna elements.
Most significantly, the intensity rapidly diminishes for angles below $20^\circ$ and above $160^\circ$.
This effect indicates a small mismatch of the indices of refractivity above and below dipole.
The angular dependence is consistent with that predicted for a ratio of about $1.52/1.44=1.06$, corresponding to the indices of refractivity of the substrate and the SiO$_x$ layer, as shown by the open circles in the figure.

In order to confirm that the reflector and director elements have the desired effect on the directionality of the emission, we fabricated antennas composed of only two elements, the feed and either a reflector or a director.
Figure 3a shows the measured emission pattern of the feed-reflector antenna (solid dots). Clearly, the presence of the reflector 125 nm behind the feed strongly suppresses the backward emission. However, the angle of emission is still rather wide. 
Figure 3b shows the measured emission pattern of the feed-director antenna (solid dots). 
Note that the spacing of 125 nm between the elements is the same as that for the feed-reflector antenna, but the director is now in front of the feed. 
The results show how the capacitive detuning of the director enhances emission in the direction of the passive antenna element. 
The angle of emission in the forward direction is even somewhat narrower than that of the feed-reflector antenna.
Both results are in general agreement with the coupled dipole predictions \cite{OursNJP07} shown by the open circles in Figs.\ 3a and 3b, confirming that the basic principles of radio frequency Yagi-Uda antennas can be applied to nanorods operating in the optical regime.

The results for a complete five element Yagi-Uda antenna are shown in Fig.\ 4.
As can be seen in the  scanning electron microscope (SEM) image in the inset, this antenna array consists of a feed, a reflector and three directors, with distances of 125 nm between the feed and the reflector and 150 nm between the directors.
According to our theoretical predictions \cite{OursNJP07}, these dimensions should be sufficiently close to the optimal distances of $\lambda/(4 n)=114$ nm and $\lambda/(\pi n)=145$ nm for a medium with an index of refractivity of $n \approx 1.45$.
The radiation pattern obtained from this antenna array shows a directionality with the emission concentrated around angles close to the region around $0^\circ$ to $20^\circ$, where the small mismatch of the index of refractivity suppresses the emission.
Taking into account this suppression of a significant part of the forward emission, we observe fairly good qualitative agreement between the experimental results (solid dots) and the theoretical prediction from the dipole model (circles).
In particular, the experimental data clearly shows a side lobe around $70^\circ$ and a small backward emission separated by a minimum at $110^\circ$.
These features of the emission pattern originate from the characteristic interferences between radiation from different array elements, indicating that the directionality of the emission is indeed obtained from the coupling of the array elements.

In conclusion, we have experimentally demonstrated the realization of a nano-optical Yagi-Uda antenna composed of an array of appropriately tuned gold nanorods. 
Our results clearly show that the basic principles of radio frequency antenna design can be applied successfully to obtain the same kind of directivity in the optical regime.
It is therefore possible to guide the emission and absorption of light using metallic structures with a total size that is not much larger than a single wavelength.
As the results for two element antennas show, even very simple arrangements can significantly improve the directionality of emission. Thus, nano-optical antenna arrays have the potential of becoming a highly flexible and useful component of future nano-optical technologies.

\textbf{Acknowledgements}

Part of this work was supported by the Grant-in-Aid program of the
Japan Society for the Promotion of Science, JSPS.

\end{document}